\documentclass[aps,prl,reprint,twocolumn,amsmath,amssymb,citeautoscript]{revtex4-1}
\usepackage{graphicx}
\usepackage{dcolumn}
\usepackage{bm}
\usepackage{latexsym,epsfig}
\usepackage{graphicx}
\usepackage{verbatim}
\usepackage{comment}
\usepackage{amsmath}
\usepackage{amssymb}
\usepackage{color}
\usepackage{epstopdf}
\usepackage{grffile}
\usepackage{lipsum}
\usepackage{enumitem}
\usepackage{subfigure}

\DeclareGraphicsExtensions{.eps}

\newcommand{\beq}{\begin{equation}}
\newcommand{\eeq}{\end{equation}}
\newcommand{\bea}{\begin{eqnarray}}
\newcommand{\eea}{\end{eqnarray}}
\newcommand{\ben}{\begin{eqnarray*}}
\newcommand{\een}{\end{eqnarray*}}
\newcommand{\bfig}{\begin{figure}}
\newcommand{\efig}{\end{figure}}

\usepackage{hyperref}
\hypersetup{
    colorlinks=true,      
    urlcolor=blue,
    citecolor=blue,
    linkcolor=blue
}

\begin{document}
\title{Re-entrant localization transition in a quasiperiodic chain}
\author{Shilpi Roy$^1$, Tapan Mishra$^1$, B. Tanatar$^2$, Saurabh Basu$^1$}
\affiliation{$^1$ Department of Physics, Indian Institute of Technology Guwahati-Guwahati, 781039 Assam, India}
\affiliation{$^2$ Department of Physics, Bilkent University, TR-06800 Bilkent, Ankara, Turkey}
\date{\today}

\begin{abstract}
Systems with quasiperiodic disorder are known to exhibit localization transition in low dimension. After a critical strength of 
disorder all the states of the system become localized, thereby ceasing the particle motion in the system. 
However, in our analysis we show that in a one dimensional dimerized lattice with staggered quasiperiodic disorder, after the localization 
transition some of the localized eigenstates become extended for a range of intermediate disorder strengths. 
Eventually, the system undergoes a second localization transition at a higher disorder strength leading to all states localized. 
We also show that the two localization transitions are associated with the mobility regions hosting the 
single particle mobility edges. We establish this re-entrant localization transition by analyzing the eigenspectra, 
participation ratios and the density of states of the system.

\end{abstract}

\maketitle

{\em Introduction.-}
The phenomenon of localization of quantum particles which is directly related to the transport properties 
has been a topic of paramount interest in recent years~\cite{Lee}. 
Originally proposed in the context of condensed matter 
systems, this phenomenon deals with the localization 
of the single particle wave function in presence of uncorrelated (random) disorder known as the Anderson localization (AL)~\cite{Anderson}.
Anderson localization predicts the metal-insulator transition as a result of quantum interference of 
scattered wave function in presence of impurities in the system. This interesting phenomenon has been studied 
in disparate systems such as photonics lattices, elastic media as well as in optical 
lattices~\cite{schwartz2007,hu2008,Drese1997,semeghini2015measurement,jendrzejewski2012three, McGehee, Roati, Billy}.

While the metal-insulator transition associated with the AL is limited to higher dimensional systems, similar 
physics can be obtained in one dimension 
by replacing the uncorrelated (random) disorder by a quasiperiodic potential. The simplest example of such 
quasiperiodic systems which are neither periodic nor completely disordered is the self-dual Aubry-Andr\'e model (AA)\cite{Aubry} which 
exhibits the localization transition at a critical quasiperiodic potential strength before (after) which all the 
states of the system are extended (localized). However, in certain generalized AA model and other quasiperiodic 
systems ~\cite{Das_Sarma_1986, Biddle, Ganeshan,Sun_2015, Gopalakrishnan,DharAA}
the transitions to the localized phases are often associated with 
a critical region where both extended and localized states coexist. The key feature of this critical region 
is the existence of the single particle mobility edge (SPME) which corresponds to a critical 
energy separating the extended and the localized states of the system\cite{Lee, Mott_1987}. 
Due to the recent progress in the field of quantum gases in optical lattices, the 
localization transition and the possible existence of the SPME in quasiperiodic systems have gained considerable 
interest \cite{Boers, Li} leading to their recent experimental observations~\cite{Luschen,An,Gadway2020} .

So far it has been well established that after the system undergoes a localization transition, all 
the states remain localized forever with increasing disorder strength. In this work we show that this is indeed not always true. In a 
one dimensional dimerized lattice with staggered quasiperiodic disorder, the competition between dimerization and quasiperiodic disorder 
leads to a re-entrant localization transition. This means, some of the already localized states become extended again 
for a range of quasiperiodic potential. Further increase in the disorder strength leads to 
the second localization transition where all the states become localized again. This re-entrant localization transition is 
also associated with separate critical regions hosting the SPMEs in the spectrum. 

{\em Model and approach.-}
We consider a one dimensional dimerized lattice with onsite quasiperiodic disorder which is given by the Hamiltonian; 
\begin{align}
H={}&-t_{1}\sum_{i=1}^{N}(c^{\dagger}_{i,B}c_{i, A}+{\rm {H.c.}})\nonumber
-t_{2}\sum_{i=1}^{N-1}(c^{\dagger}_{i+1, A}c_{i, B}+ {\rm {H.c.}})\\ 
&+\sum_{i=1}^{N} \lambda_{A}n_{i,A}\cos[2\pi\beta(2i-1)]+\sum_{i=1}^{N}\lambda_{B}n_{i,B}\cos[2\pi\beta(2i)]
\label{eqn:ham}
\end{align}
This is a chain of $N$ unit cells consisting of two sublattice sites $A$ and $B$. 
$i$ represents the unit cell index and $L=2N$ is the length of the chain. $ c^{\dagger}_{i,A}~(c_{i, A})$ 
and $c^{\dagger}_{i, B}~(c_{i, B})$ are the creation (annihilation) operators 
corresponding to sites in the $A$ and $B$ sublattices which we denote by ($i, A$) and ($i, B$) 
and the site number operators are 
denoted as $n_{i,A}$ and $n_{i,B}$. 
The intra- and inter-cell hopping strengths are represented by $t_{1}$ and $t_{2}$ respectively and H.c. stands of the Hermitian conjugate. 
The strength of the onsite potential at the sublattice site $A~(B)$ is represented by $\lambda_{A}$~($\lambda_B$) and 
$\beta$ determines the period of quasiperiodic potential. The staggered disorder is introduced 
by assuming $\lambda_A=-\lambda_B=\lambda$ in Eq.~\ref{eqn:ham}. 
The model Eq.~\ref{eqn:ham} in the limit of vanishing disorder i.e. $\lambda=0$,
is the paradigmatic Su-Schrieffer-Heeger (SSH) model \cite{Su} which exhibits a trivial (when $t_1 > t_2$) to topological 
( when $t_1 < t_2$) phase transition through a gap closing point at $t_1=t_2$. This phase transition is 
protected by the chiral symmetry of the system. Note that in presence of finite onsite disorder this symmetry is explicitly broken. 

We choose $\beta=(\sqrt{5}-1)/2$, a Diophantine number \cite{Jitomirskaya} in our work and fix the 
intra-cell hopping, $t_1=1$ as the energy scale. 
For convenience, we define a quantity $\delta=t_2/t_1$ which controls the hopping dimerization in Eq.~\ref{eqn:ham}. 
The system size considered in our simulations is up to $L=13530$, that is, $N=6765$ unit cells. We explore the effect of staggered disorder in both the 
limits of dimerization in Eq.~\ref{eqn:ham} such as $\delta <1$ and $\delta >1$) . 
To analyze the physics of the model shown in Eq.~\ref{eqn:ham},  we rely on the inverse participation ratio (IPR) and the 
normalized participation ratio (NPR) \cite{Li1,Li2}, which are the two most significant diagnostic tools to characterize 
the localization transition.  For the $n$-th eigenstate, $\phi^{i}_{n}$, the IPR and the NPR are defined as,
\begin{equation}
{\rm{IPR}}_{n}=\sum_{i=1}^{L} |\phi^{i}_{n}|^{4} ~~~,~~~~~~ {\rm{NPR}}_{n}=\bigg(L\sum_{i=1}^{L} |\phi^{i}_{n}|^{4}\bigg)^{-1}.
\end{equation}
The extended(localized) phases are characterized by $\rm {IPR}=0(\neq 0)$ and $\rm{NPR}\neq 0(=0)$ in the large $L$ limit.  
Before proceeding with the staggered $\lambda$ case we 
first highlight the physics associated with the case of uniform $\lambda$ for comparison.

\begin{figure}[t]
\centering
{\includegraphics[width=3.45in]{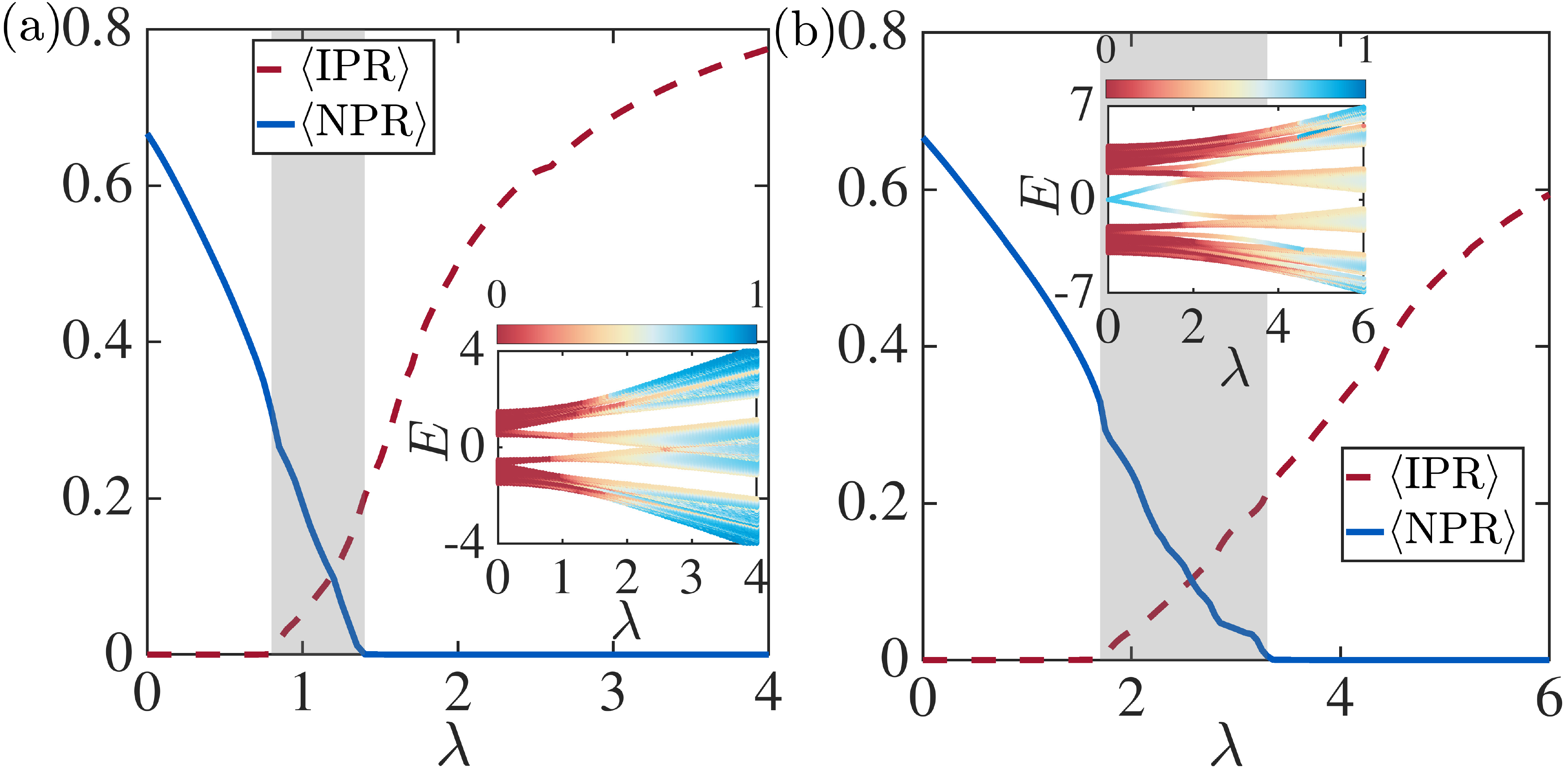}}
\caption{Figure shows the $\langle \rm{IPR} \rangle$ (red-dashed) and $\langle \rm{NPR} \rangle$ (blue-solid) 
are plotted as a function of $\lambda$ for (a) $\delta=0.5$ and (b) $\delta=3$ for a system of size $L=13530$. 
The shaded regions indicate the critical or the intermediate regimes. 
The color maps in the insets show the plots of IPR associated to all eigenmodes $E$ with respect to $\lambda$ for values of $\delta$ of the main figure. }
\label{fig:IPR_NPR_equal}
\end{figure}

{\em Uniform disorder} ($\lambda_A=\lambda_B=\lambda$).- 
In the limit of $\delta=1$, Eq.~\ref{eqn:ham} corresponds to the pure AA model which exhibits a localization transition 
without any SPME. However, moving away from this limit, we show that the localization transition 
occurs through a critical regime hosting the SPME for both $\delta <1$ and $\delta > 1$. 
To identify the localization transitions we plot the $\langle \rm{IPR}\rangle$  and the $\langle \rm{NPR}\rangle$ 
as a function of $\lambda$ 
for the two exemplary points, namely, $\delta= 0.5$ and $\delta= 3$ in Fig.~\ref{fig:IPR_NPR_equal}(a) and (b) respectively. 
Here $\langle \rm{IPR}\rangle$  and $\langle \rm{NPR}\rangle$ denote the averages of the 
IPR and NPR computed by considering all the eigenstates for a particular value of $\lambda$. 
It can be seen that contrary to the simple AA model ($\delta=1$), 
the plots for the $\langle \rm{IPR}\rangle$ and the $\langle \rm{NPR}\rangle$ do not sharply 
cross each other at the duality point $\lambda=2$\cite{SOKOLOFF1985} for both the values of $\delta$. Rather they 
cross each other at very different values of $\lambda$ and also exhibit a coexisting 
region where both the $\langle \rm{IPR}\rangle$ 
and  the $\langle \rm{NPR}\rangle$ are finite (shaded regions). This signifies the presence of both the localized and the 
extended states for a range of $\lambda$ ($0.7 < \lambda < 1.4$ when $\delta=0.5$ 
and $1.6 < \lambda < 3.4$ when $\delta=3$) which are the critical phases exhibiting the SPMEs. 
Clearly, after the localization 
transition all the states remain localized as a function of $\lambda$.

The localization transition and the existence of the SPME can be easily inferred from the energy 
spectrum and the associated IPR of the individual states. 
We plot the IPR associated to the energy spectra, $E$ corresponding to the Hamiltonian in Eq.~\ref{eqn:ham} 
for $\delta=0.5$ and $3$  in the insets of Fig.~\ref{fig:IPR_NPR_equal}(a) and (b) respectively. 
Here, the eigenenergies are color coded with the corresponding IPRs. 
Due to the dimerized nature of the model (Eq.~\ref{eqn:ham}), we get two distinct 
energy bands at $\lambda=0$ and in this limit the energy levels are completely 
extended for both the trivial (Fig.~\ref{fig:IPR_NPR_equal}(a)) 
and the topological (Fig.~\ref{fig:IPR_NPR_equal}(b)) cases. 
As the value of $\lambda$ increases, the gaps between the bands in both the dimerized 
limits tend to vanish beyond a critical $\lambda$. 
Clearly, in both the cases, the fully extended (red) and the localized regions (blue) are separated by a critical phase where both extended and localized states coexist for a range of values of $\lambda$ which host a SPME. 
Quite expectedly,  the appearance of the localized states at $\lambda=0$ in the inset of Fig.~\ref{fig:IPR_NPR_equal}(b), are the 
topological edge modes present in the middle of the gap. 
We shall discuss the fate of these edge modes later.
Note that other minibands with some states in the gaps between 
them appear in the energy spectrum due to the quasiperiodic disorder which are irrelevant for the present analysis.

{\em Staggered disorder} ($\lambda_A=-\lambda_B=\lambda$).-
In this section we discuss the role of staggered disorder on the localization transition. 
In this case also one expects a qualitatively similar localization transition as in the 
uniform disorder case with some quantitative difference. 
This is confirmed in our analysis which shows the extended to localization transition through a critical phase 
where both  $\langle \rm{IPR}\rangle$  and $\langle \rm{NPR}\rangle$ are finite for a range of values of $\delta$. 
As it is well known and already mentioned before, in quasiperiodic lattices exhibiting localization hosting the SPME,  
for the values of $\lambda$ prior to (beyond) the critical phase, all the states of 
the system are extended (localized). 
\begin{figure}[t]
\centering
{\includegraphics[width=3.45in]{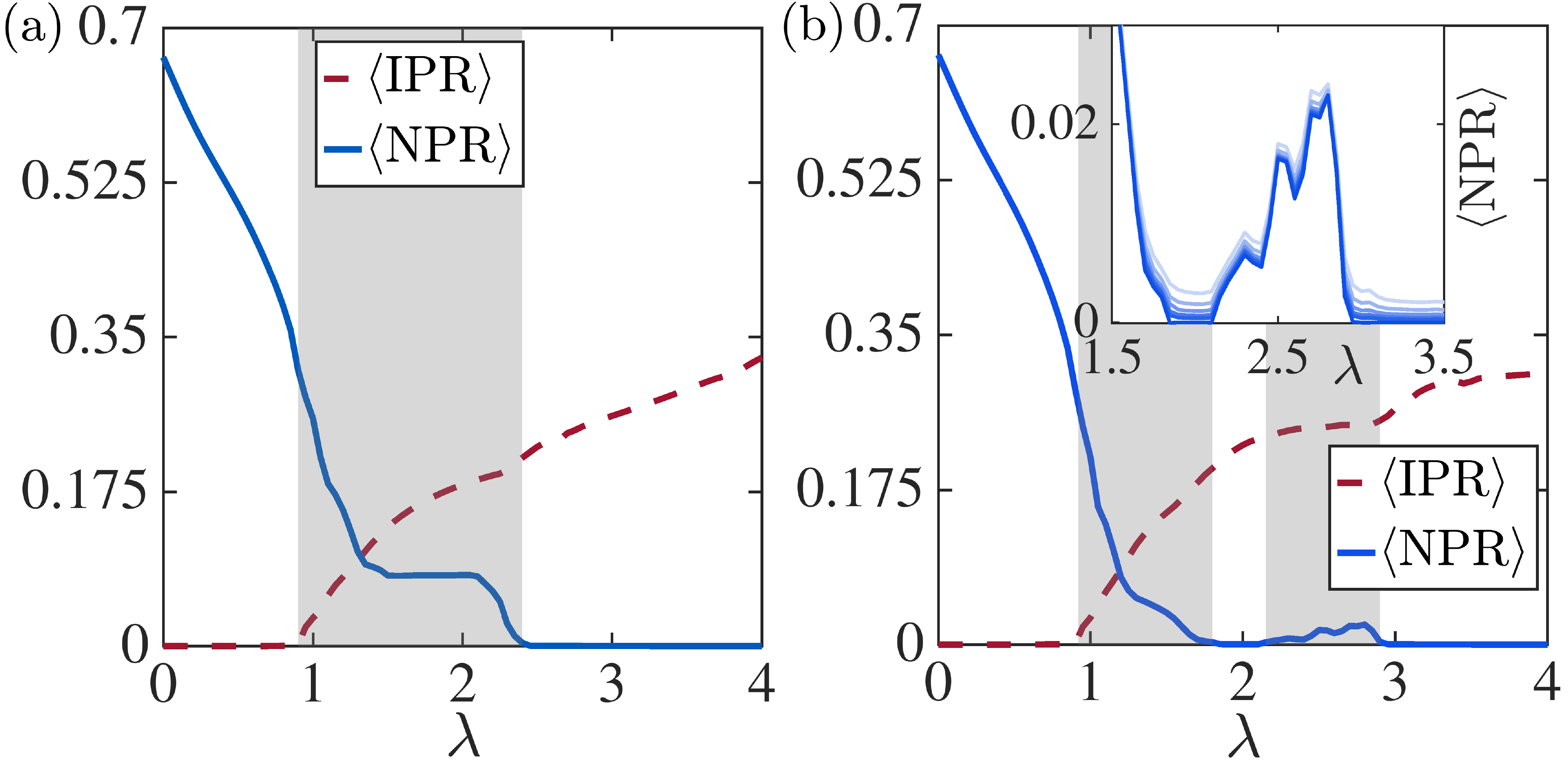}}
\caption{(a) and (b) show the $\langle \rm{IPR}\rangle$ and the 
$\langle \rm{NPR}\rangle$ for $\delta=1.5$ and $2.2$ respectively for the case of staggered disorder and $L=13530$. The shaded regions represent 
the critical phases. (Inset) shows the $\langle \rm{NPR}\rangle$ for $L=1974,~3194,~5168,~8362,~13530~{\rm and}~\infty$ (light to deep blue).}
\label{fig:IPR_NPR_opp}
\end{figure}

\begin{figure}[b]
\centering
{\includegraphics[width=3.45in]{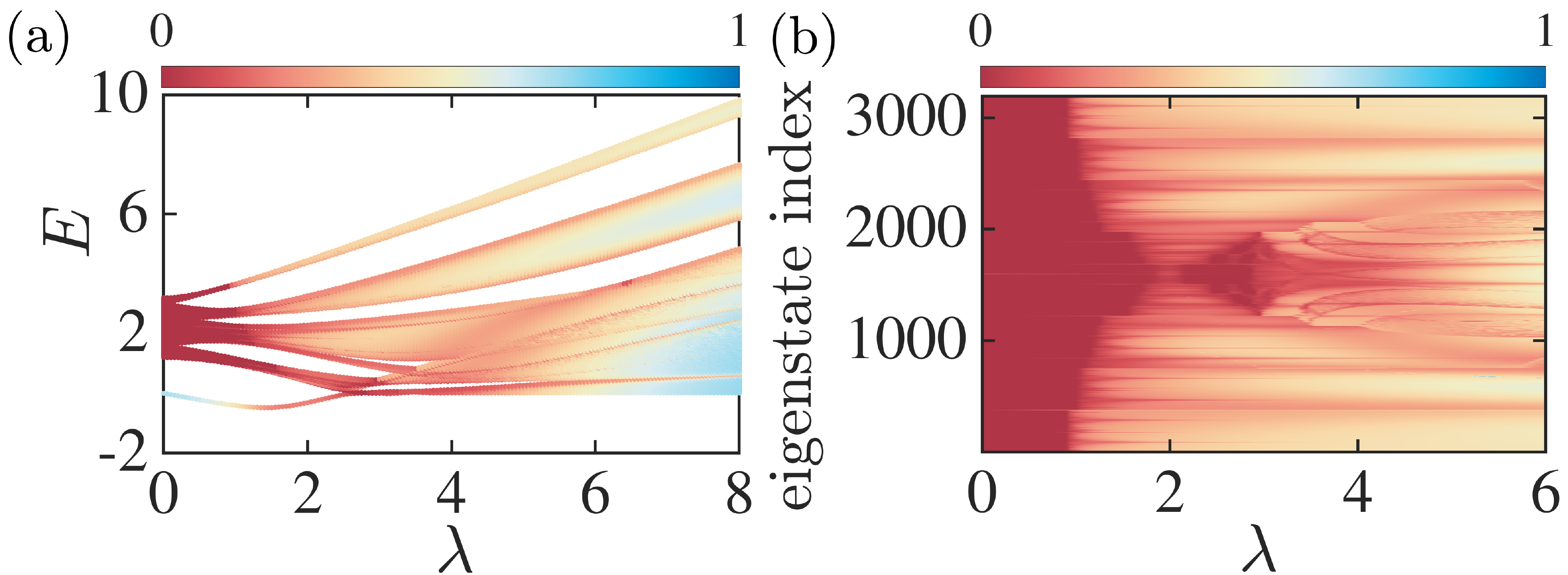}}
\caption{(a) The upper half of the energy eigenvalue spectrum superimposed with their respective IPR shows the extended, critical and localized states. 
(b)The IPR associated to the eigenstate indices as a function $\lambda$ for $\delta=2.2$ for a system 
of size $L=3194$. }
\label{fig:ener_ipr}
\end{figure}

\begin{figure}[t]
\centering
{\includegraphics[width=\columnwidth]{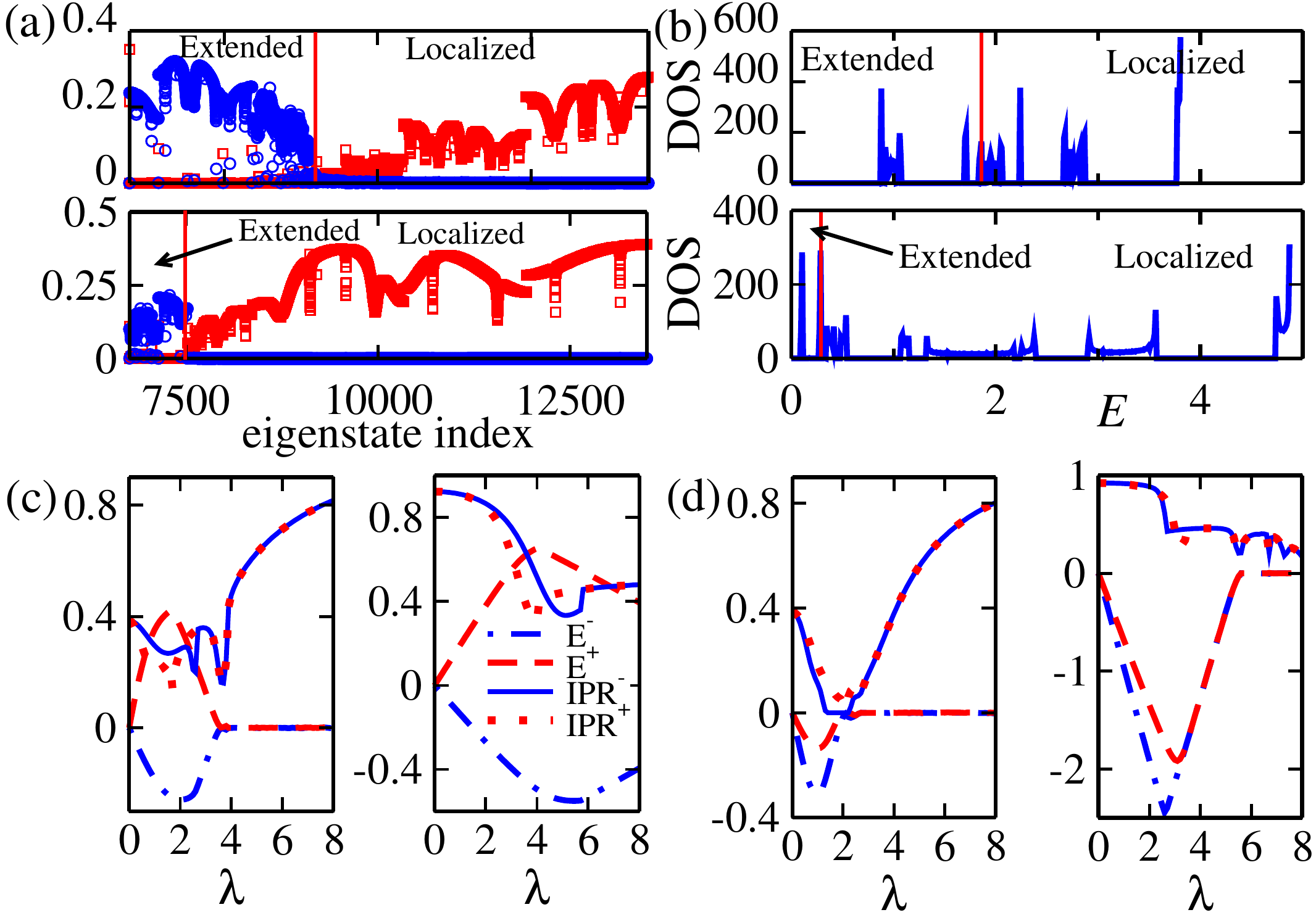}}
\caption{(a) Shows the IPR (red squares) and NPR (blue circles) of 
different eigenstates  for $\delta=2.2$ and $\lambda=1.2$ (upper panel) and $\lambda=2.5$ (lower panel). 
The states with finite IPR in the extended regime are the emerging 
edge modes in the fractal gaps. (b) Shows the DOS for  $\delta=2.2$ and $\lambda=1.2$ (upper panel) 
and $\lambda=2.5$ (lower panel). The vertical lines separate the extended and localized regions.
(c) and (d) show the edge states and the corresponding IPRs for uniform and staggered disorder respectively for 
$\delta=1.5$ (left panel) and $\delta=5$ (right panel). $E^-$ (blue dot-dashed) and $E^+$ (red dashed)
corresponding to the two edge states along with their IPR i.e. IPR$^-$(blue solid) and IPR$^+$ (red dotted). For better visibility the energies in the right panel of (c) are plotted as $E/4$. }
\label{fig:Edge_IPR}
\end{figure}
Once the system is in the localized phase, it remains localized as a function of the strength of the quasiperiodic potential, $\lambda$. 
As a result, one gets $\langle \rm{IPR}\rangle \neq 0$ and $\langle \rm{NPR}\rangle = 0$ for all values of $\lambda$ 
after the critical regime. 
However, surprisingly in presence of the staggered disorder, we find that for some intermediate values of $\delta$,  
the system undergoes two localization transitions through two critical phases as a function of $\lambda$. 
This re-entrant localization behaviour can be very well discerned 
by together analyzing the $\langle \rm{IPR}\rangle$ and $\langle \rm{NPR}\rangle$. 
In Figs.~\ref{fig:IPR_NPR_opp}(a) and (b) we show the $\langle \rm{IPR}\rangle$  
and $\langle \rm{NPR}\rangle$ corresponding to two different values of dimerization such as $\delta=1.5$ and $2.2$ respectively for $L=13530$.  
Clearly for $\delta=1.5$ (Fig.~\ref{fig:IPR_NPR_opp}(a)), there is a 
transition to the localized phase through a critical region for the range of $\lambda$ between $0.9 < \lambda < 2.5$. After the localization 
transition i.e. for $\lambda > 2.5$, all the states are localized. 
On the other hand, for $\delta=2.2$ (Fig.~\ref{fig:IPR_NPR_opp}(b)), there exists two 
critical regions in the range $0.9 < \lambda < 1.8$ and $2.1 < \lambda < 2.9$ where both the $\langle \rm{IPR}\rangle$  and 
$\langle \rm{NPR}\rangle$ are finite. 
In the region between the  two critical phases and again beyond the second critical phase, the system is fully localized. This 
indicates that the system also hosts two SPMEs as a function of $\lambda$. Note that the extent of the second 
critical region occurs for a small range of $\lambda$ . In order to rule out any finite size effects, we perform finite size extrapolation of the $\langle \rm{IPR}\rangle$~\cite{Supmat} and $\langle \rm{NPR}\rangle$ considering data for different system sizes such as $L=1974,~ 3194,~5168, ~8362,~13530$. The inset of 
Fig.~\ref{fig:IPR_NPR_opp}(b) shows the $\langle \rm{NPR}\rangle$ data for various system sizes including the one at $L\to\infty$ for $\delta=2.2$. This clearly indicates the stability of the second critical region. 

This re-entrant localization feature can be seen in the energy spectrum encoded with the corresponding IPR 
as shown in the Fig.~\ref{fig:ener_ipr}(a). 
For clarity we depict only the upper band of the spectrum which shows a transition from extended-critical-localized-critical-localized regions as 
a function of $\lambda$. A clear picture can be obtained by plotting the IPR of the individual eigenstates as shown 
in Fig.~\ref{fig:ener_ipr}(b). The deep-red patches appearing in Fig.~\ref{fig:ener_ipr}(a) and (b) for $2.1 < \lambda < 2.9$ indicate that some of the localized states become extended again for a range of $\lambda$. This confirms the presence of the second critical region and the second SPME. 
We further confirm the existence of the SPME by analyzing the IPR and the NPR for the individual eigenstates 
of the system in the critical regime. Figure~\ref{fig:Edge_IPR}(a), shows the IPR and NPR for all the eigenmodes for 
$\delta=2.2$ at $\lambda=1.2$ and $2.5$ in the upper and lower panels respectively. 
The plots show a clear distinction between the extended states (finite NPR) from the 
localized states (finite IPR) of the spectrum. Similar signature is also seen in the density of states (DOS)
by analyzing it with the IPR of the individual states indicating the existence of the mobility 
edge as shown in Fig.~\ref{fig:Edge_IPR}(b) (see figure caption for detail). 



\begin{figure}[t]
\centering
{\includegraphics[width=\columnwidth]{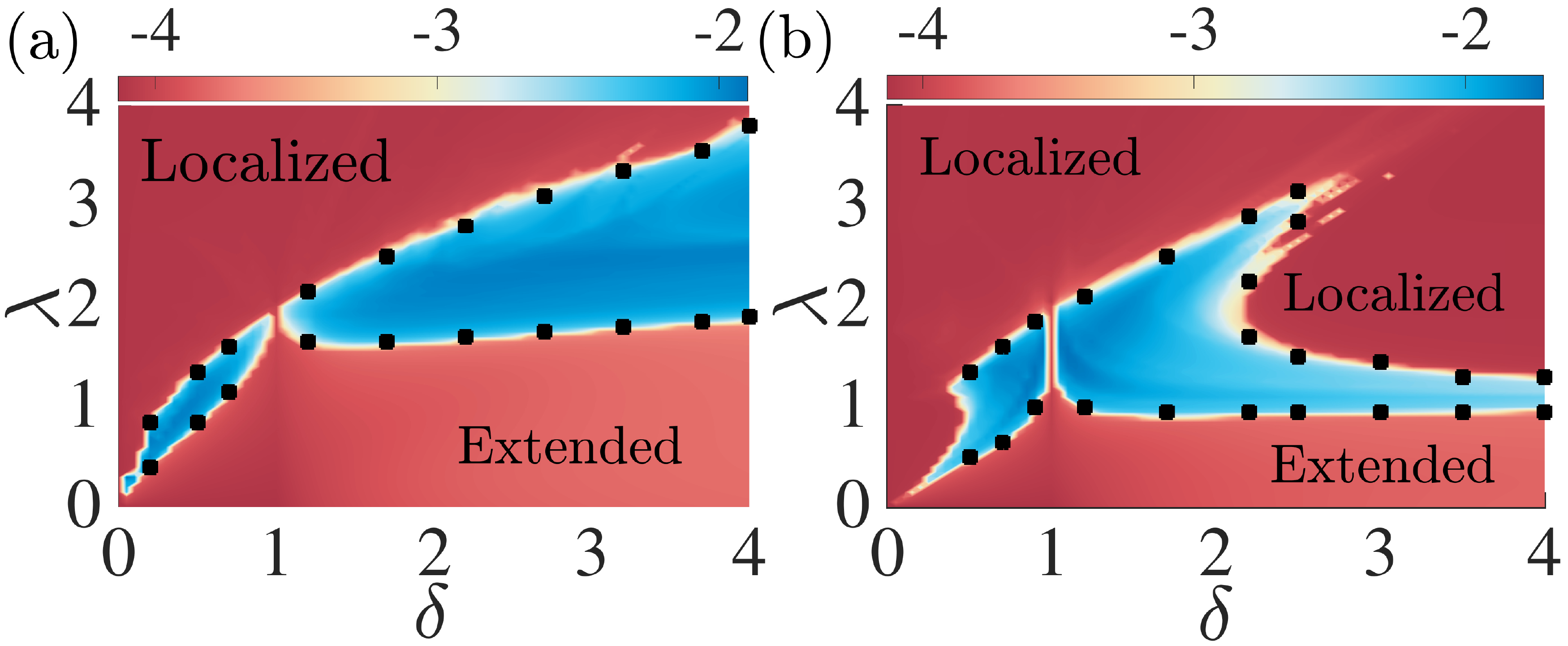}}
\caption{ The phase diagrams in $\delta$ and 
$\lambda$ plane for (a) uniform disorder and 
(b) staggered disorder cases. The filled black squares are the data points obtained by examining the $\langle \rm {IPR}\rangle$ and $\langle \rm {NPR}\rangle$ plots.
(See text for details). The color code indicates the values of $\eta$.} 
\label{fig:phase_diagram}
\end{figure}

{\em Phase diagram:-}
Finally, we present the key results in the form of a phase diagram as displayed in Fig.~\ref{fig:phase_diagram}(b) for the case of 
staggered disorder in the $\delta$ - $\lambda$ plane. The phase diagrams is obtained 
by computing a quantity $\eta$ introduced in Ref.~\cite{Li2} as;
\begin{equation}
 \eta=\rm log_{10}[\langle \rm IPR\rangle \times \langle \rm NPR\rangle]
\end{equation}
The presence of the critical region (blue region bounded by the symbols) is clearly 
distinguished from the fully extended or the fully localized 
regions (red regions) in the phase diagram. Note that the critical regions are separated by a narrow passage at $\delta=1$ (AA model),
where a sharp localization transition occurs. It can be seen that for a range of $\delta$ one encounters two critical regimes with increase in $\lambda$ which is 
the key finding of our analysis. 
However, this re-entrant feature does not appear in the case of uniform 
disorder (compare Fig.~\ref{fig:phase_diagram}(a)). 
We complement the above findings by directly locating the boundaries (filled squares) of the critical region by examining the 
values of $\langle \rm IPR\rangle$ and $\langle \rm NPR\rangle$ in the theormodynamic limit. This non-trivial feature of the 
re-entrant localization transition and the SPME can be attributed to the competition between the hopping dimerization and the 
staggered disorder that renders extended nature to some of the low energy localized states. 
The detailed analysis above requires further investigation.

It is worth mentioning that the re-entrant localization phenomena and the mobility edge occurs in both the limits of the dimerization (see Fig.~\ref{fig:phase_diagram}(b). 
Hence, an important conclusion can be drawn at this point is that the underlying topological 
properties has no role in establishing the 
re-entrant localization transition. 

{\em Edge modes.-}
Having analyzed the physics of the bulk spectrum, we discuss the fate of the topological zero energy edge modes 
as a function of the disorder strength. We note that the initially localized 
zero modes (at $\lambda=0$) become energetic and finally hybridize into the bulk bands with increase in $\lambda$ 
 for both uniform and staggered disorder cases as already shown in Fig.~\ref{fig:IPR_NPR_equal} (b) (inset) and Fig.~\ref{fig:ener_ipr}(a) respectively. 
To explicitly understand the behavior of these modes, we separately plot the edge modes as a 
function of $\lambda$ in Fig.~\ref{fig:Edge_IPR} along with their IPR. 
We consider two different values of $\delta$, namely,  
$\delta=1.5$ and $\delta=5$ which represent weak and strong dimerization limits pertaining to the topological regime. 
As mentioned earlier, owing to the breaking of the chiral symmetry induced by the quasiperiodic potential, 
both the edge modes, namely,
the particle mode ($E^+$ shown by a dashed red line) and the hole mode ($E^-$ shown by a dot-dashed blue line) 
asymmetrically separate out from each other towards the opposite 
bands as  $\lambda$ increases (Fig.~\ref{fig:Edge_IPR}(c)) for the case of uniform potential. 
However, in the case of the staggered disorder, both the edge modes move differently towards the 
lower band (Figs.~\ref{fig:Edge_IPR}(d))~\cite{Backgroundnote}.
Eventually for all the cases, beyond certain critical values of $\lambda$, $E^+$ and $E^-$ 
tend to merge with each other. 
We also plot the corresponding IPR for both the modes as IPR$^+$ (dotted red) and IPR$^-$ (solid blue). 
It can be seen that in 
all the four cases the IPR initially decreases and 
then increases as a function of $\lambda$. In the case of weak dimerization,
initially the states are localized. As the value
of $\lambda$ increases, the states become delocalized first and then become strongly localized. On the other hand, in the
case of strong dimerization , the states which are strongly localized 
(IPR$\sim 1$) at the beginning (for small $\lambda$) remain localized forever. 
This analysis indicates that the behaviour of the edge states as a function of $\lambda$ is independent of the bulk behaviour.


{\em Conclusions.-}
We have studied the localization transition in a dimerized lattice with staggered quasiperiodic disorder. We 
show that the system undergoes a re-entrant localization transition as a function of disorder strength for a range of values of dimerization. 
The re-entrant localization occurs in both the regimes of dimerization and each localization transition is associated with the SPME. 
We confirm this finding by examining the participation ratios, 
the single particle spectrum and the behavior of the individual eigenstates and present a 
phase diagram depicting all the above findings. For completeness we also analyse the phase diagram in the case of uniform disorder which 
shows the usual localization transition and the SPME. In the end we discuss the fate of the zero energy edge modes 
as a function of disorder strength which were initially localized in 
the absence of any disorder due to the topological nature of the model.

The re-entrant feature may reveal interesting physics in transport and dynamical properties of quantum particles.
An immediate extension could be to study the
stability of this re-entrant phenomenon in the context of many-body localization. 
Due to the phenomenal experimental progress in systems of ultracold atoms in 
optical lattices to simulate dimerized latticed~\cite{Lohse2015,Leseleuc2018}, 
quasiperiodic systems~\cite{Luschen,Gadway2020} and the recent experiment on disorder induced 
topological phase transition using $^{\rm{171}}$Yb~\cite{Takahashi2020},
our findings can in principle be simulated in the state-of-the art quantum gas experiments. 

{\em Note added:} While preparing the manuscript, we became aware of an interesting recent work
related to localization transition in 
an interpolating Aubry-Andr\'e-Fibonacci (IAAF) model~\cite{GoblotAA2020}. The model is shown to exhibit a 
cascade of band selective localization/delocalization 
transitions while transiting from the AA into a Fibonacci model. 

{\em Acknowledgments.-} 
{We thank Luis Santos and Subroto Mukerjee for useful discussions. 
SB acknowledges funding from Science and Engineering Research
Board (SERB) India (Project No. EMR/2015/001039). 
BT acknowledges support from TUBA and TUBITAK. TM acknowledges financial support from Science and Engineering Research
Board (SERB) India (Project No. ECR/2017/001069).}

\bibliography{refs}


\end{document}


\title{Supplementary material for ``Re-entrant localization transition in a quasiperiodic chain"}
\author{Shilpi Roy$^1$, Tapan Mishra$^1$, B. Tanatar$^2$, Saurabh Basu$^1$}
\affiliation{ Department of Physics, Indian Institute of Technology Guwahati-Guwahati, 781039 Assam, India}
\affiliation{Department of Physics, Bilkent University, TR-06800 Bilkent, Ankara, Turkey}
\date{\today}

\date{\today}

\begin{abstract}
In this Supplementary Material we provide the finite size analysis of the participation ratios to concretely establish the re-entrant localization as discussed in the main text of the paper.
\end{abstract}

\maketitle

In order to verify the existence of the re-entrant region in the thermodynamic limit we compute the $\langle \rm{IPR} \rangle$ and $\langle \rm{NPR} \rangle$ for different systems sizes such as $L=1974, 3194, 5168,
8362~ \rm{and}~ 13530$ and perform finite size extrapolation as shown below. The Fig.~\ref{fig:scalling_IPR}(a) and Fig.~\ref{fig:scalling_NPR}(a) show the values of
$\langle \rm{IPR} \rangle$ and $\langle \rm{NPR} \rangle$  respectively, plotted
against $\lambda$ for $\delta=2.2$
corresponding to different system sizes and also in the thermodynamic limit (light to deep blue
curves). This clearly indicates the existence of a second critical
region even in the large $L$ limit indicating the re-entrant behaviour
of the localization transition. 
\begin{figure}[!b]
\centering
{\includegraphics[width=3.in]{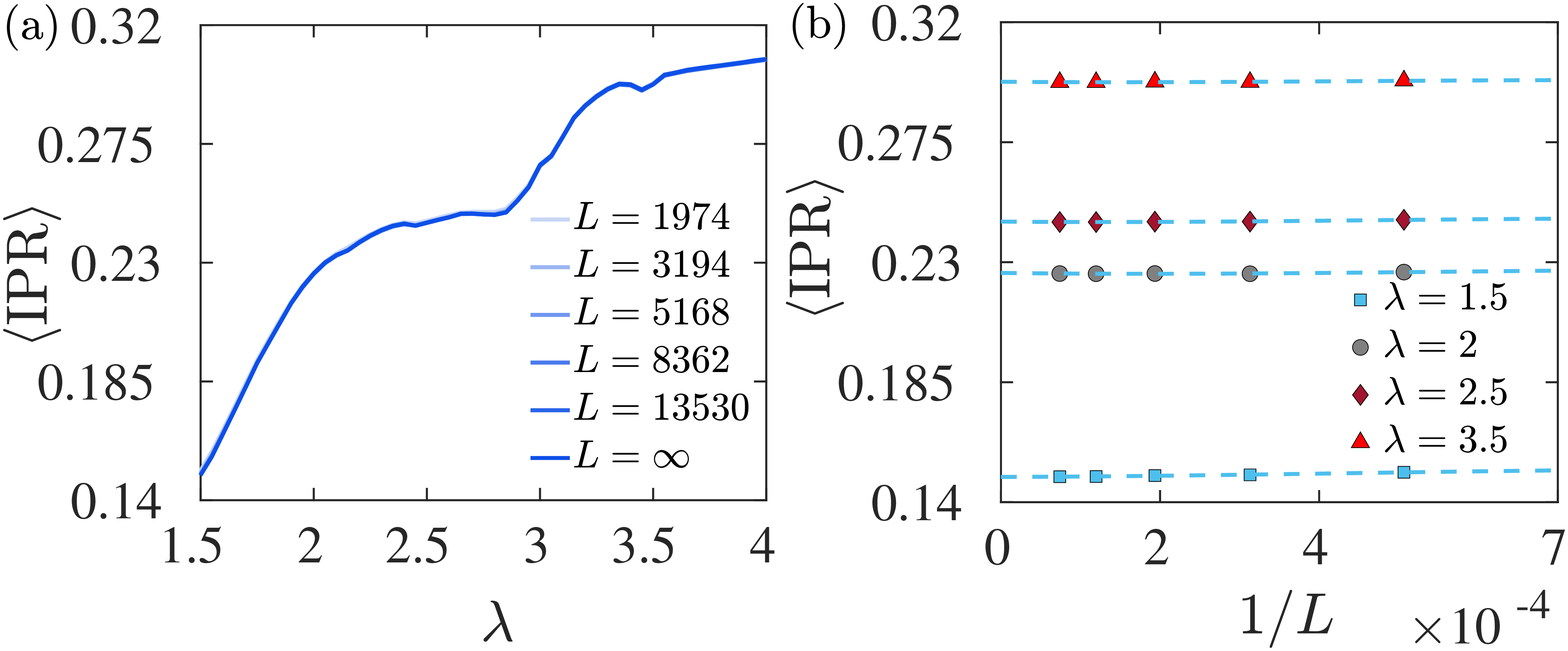}}
\caption{(a) The  $\langle \rm{IPR} \rangle$ for all the lengths including the one at the thermodynamic limit are shown which show negligible finite size effects. (b) Finite size extrapolation of the $\langle \rm{IPR} \rangle$ is shown for some selected values of $\lambda$. It can be seen that the $\langle \rm{IPR} \rangle$ remains finite throughout the region of interest. }
\label{fig:scalling_IPR}
\end{figure}

\begin{figure}[!b]
\centering
{\includegraphics[width=3.2in]{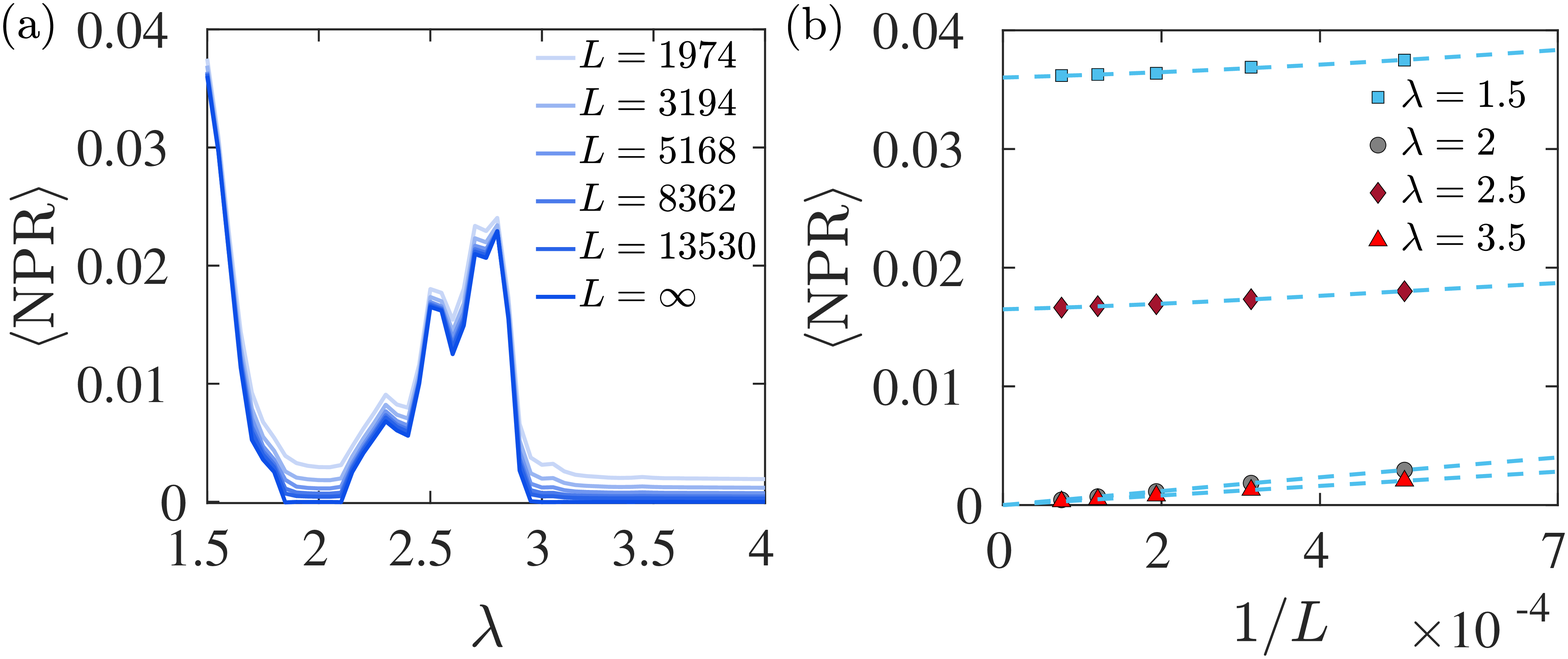}}
\caption{(a) The  $\langle \rm{NPR} \rangle$ for all the lengths including the one at the thermodynamic limit clearly shows the re-entrant behaviour.  (b) Finite size extrapolation of the $\langle \rm{NPR} \rangle$ is shown for some selected values of $\lambda$.  }
\label{fig:scalling_NPR}
\end{figure}
The finite size analysis $\langle \rm{IPR} \rangle$ and $\langle \rm{NPR} \rangle$ is shown in Fig.~\ref{fig:scalling_IPR}(b) and Fig.~\ref{fig:scalling_NPR}(b) respectively for some representative values of
$\lambda$ such as $\lambda=1.5,~2.0,~2.5~ {\rm and} ~3.5$. The values of $\langle \rm{IPR} \rangle$ and $\langle \rm{NPR} \rangle$ are plotted against $1/L$ and the finite size extrapolation is done by
fitting a cubic polynomial to the data points and extrapolating it to $L\to \infty$. It can be clearly seen that while the scaling of $\langle \rm{IPR} \rangle$ results in finite values throughout the region of interest, the $\langle \rm{NPR} \rangle$ clearly indicates the re-entrant transition.